\documentclass{PoS}

\title{Accreting White Dwarfs}

\ShortTitle{Accreting White Dwarfs}

\author{\speaker{Margarita Hernanz}
\\      
Institut de Ci\`encies de l'Espai (CSIC-IEEC), Bellaterra (Barcelona) and  
Institut d'Estudis Espacials de Catalunya, Barcelona, Spain\\
E-mail: \email{hernanz@ieec.uab.es}}

\author{Jordi Jos\'e\\
Departament de F\'{\i}sica i Enginyeria Nuclear, EUETIB, UPC and 
Institut d'Estudis Espacials de Catalunya, Barcelona, Spain\\
E-mail: \email{jjose@ieec.fcr.es}}

\abstract{Thermonuclear (type Ia) supernovae are explosions in accreting 
white dwarfs,
but the exact scenario leading to these explosions is still unclear. An
important step to clarify this point is to understand the behaviour of
accreting white dwarfs in close binary systems. The characteristics of the
white dwarf (mass, chemical composition, luminosity), the accreted material 
(chemical composition) and those related with the properties of the binary 
system (mass accretion rate), are crucial for the further evolution towards the
explosion. An analysis of the outcome of accretion and the implications for
the growth of the white dwarf towards the Chandrasekhar mass and its
thermonuclear explosion is presented.
}

\FullConference{Supernovae: lights in the darkness (XXIII Trobades 
Cient\'ifiques de la Mediterr\`ania)\\
		 October 3-5 2007\\
		 Mao, Menorca, Spain}

\begin{document}

\section{Introduction}
White dwarfs are the final stages of the evolution of stars with masses 
smaller than 8-11 M$_\odot$. The final fate of such stars is just cooling 
to invisibility, whenever they are isolated. However, when they belong to 
interacting binary systems, more exciting phenomena can happen, like 
nova explosions and supernovae of the thermonuclear class, i.e., type Ia 
supernova 
explosions. Accretion of matter from the companion star is at the origin of 
such interesting phenomena, but there is a long and complicated path from the 
onset of accretion up to the final explosion, in case this is the final 
outcome of the evolution.

It is widely accepted that thermonuclear supernovae are the result of the 
explosion of a carbon-oxygen (CO) white dwarf, once it reaches the critical 
density for carbon ignition; this occurs either at the center of the star 
or off-center, always in strongly degenerate conditions. Therefore, it is 
expected that the properties of such explosions are quite uniform, in 
agreement with the observations, allowing the use of them as standard candles 
for cosmological purposes. In contrast, core colllapse supernovae, where a 
massive star explodes because of the gravitational force, show a wide range 
of observational properties, since the range of masses of the progenitor 
stars is extremely large. 

In spite of the importance of type Ia supernovae both as cosmological tools, 
as crucial contributors to the chemical and dynamical evolution of the 
Galaxy, and as the triggering mechanism for star formation, their exact 
scenario is still unknown. In addition, there are not yet successful 
simulations of the explosion (see the review by \cite{HN00}). In this paper 
we concentrate on the scenario issue, with a special emphasis on the 
fundamental role played by accretion, and its link to the final fate of 
the white dwarf. The type of material accreted (either H, He -with some
metals- or CO) and the rate at which matter falls onto the white dwarf
(mass accretion rate) are strongly dependent on the type of interacting binary 
hosting the potential exploding white dwarf. The mass and chemical 
composition of the white dwarf itself, as well as its initial luminosity, 
are also crucial for its outcome. There is a complicated interplay between 
type of accretion and initial white dwarf conditions, such that it is not 
straightforward to reach the carbon ignition conditions. Once ignition occurs, 
other problems occur, related to the flame propagation; these are out of the 
scope of this paper, but are treated in other papers of this volume (e.g., 
Roepke's contribution \cite{Roepke}).

\section{Scenarios of SNIa}

There are two basic scenarios for the progenitors of type Ia supernova 
explosions: the single degenerate scenario, where the white dwarf is 
accreting matter from a main sequence or giant companion, and the double 
degenerate scenario, where the merging of two white dwarfs is the responsible 
for the final explosion. In the first case, accreted matter is always 
hydrogen-rich, except when the companion is a helium star \cite{LT91} and 
therefore accreted matter is pure helium. In the double degenerate scenario, 
accreted matter can be either pure He or a mixture of carbon and oxygen, 
depending on the type of white dwarf companion. 

The influence of accretion on the final outcome of the white dwarf 
will be discussed in next section. Here we just 
mention the observational problem related with accretion of H-rich matter, 
since it is a crucial point to disentangle between the single degenerate and 
the double degenerate scenarios.
Accretion of H-rich material poses some problems, because hydrogen is not 
seen in the spectra of type Ia supernovae (absence of hydrogen is one of 
the basic properties defining the type I -and in particular Ia- class). 
However, according to models, some hydrogen should be stripped from 
the secondary star during the explosion -in the single degenerate scenario- 
and, therefore, some hydrogen should exist in the ejecta 
\cite{Marietta00,Meng07}. This hydrogen is predicted to be at small 
velocities (lower than $\sim 10000$ km/s), and its detection is hard at the 
stages where bulk material expands at larger speeds. 
In the last years, there have been 
serious attempts to search for hydrogen in the spectra of type Ia supernovae, 
and upper limits have been obtained which do not completely contradict the
single degenerate scenario (see for instance \cite{Leonard07} and references 
therein). In fact, there has been detection of circumstellar material, 
indicative of a giant companion (i.e., single degenerate scenario) in a 
few cases \cite{Hamuy03,Patat07}.

Another issue is to know the properties of the white dwarf itself, e.g., 
mass and chemical 
composition, which are strongly related. Three compositions are possible for 
white dwarfs: helium, carbon-oxygen (CO) and oxygen-neon (ONe). The mass 
range over which the white dwarf progenitor star undergoes either the AGB 
(from Asymptotic Giant Branch), without carbon ignition and leaving CO cores, 
or the super-AGB phase, which burns carbon and leaves ONe cores, is a bit 
controversial. The mass range
for which an ONe white dwarf can form is between $\sim 8-10$ M$_\odot$, 
depending on the treatement of convection (and specially on the inclusion 
or not of overshooting) and mass loss during the previous binary evolution 
and during the thermal pulses themselves \cite{Ritossa96,ET04}. Only CO 
white dwarfs are expected to be able to explode as type Ia supernovae, since 
they are carbon-rich. On the contrary, ONe white dwarfs are expected to 
collapse, because of the effect of electron captures on $^{24}$Mg (see for 
instance \cite{Nom87,NK91,Can92,Guti06} and references therein).

Two possibilities have been suggested, concerning the mass of the 
white dwarf star: Chandrasekhar and sub-Chandrasekhar scenarios. The 
standard scenario is based on the explosion of a Chandrasekhar mass CO white 
dwarf, but it has been claimed that relatively low-mass CO white dwarfs 
could also explode. In this case, helium detonation on top of the white dwarf 
would drive the final central or off-center carbon ignition responsible 
for the explosion (see for instance \cite{WW94}). The propagation of the 
outward carbon detonation through
material at a lower density (than in the Chandrasekhar mass white dwarf)
alleviates the problem of the absence of intermediate-mass elements
(such as Si and S, for instance) of the
old pure carbon detonation models of Chandrasekhar mass white dwarfs
(see for instance \cite{Nomoto82b}).
But other problems arise in this sub-Chandrasekhar scenario, since it does
not well reproduce the observed velocities of the intermediate-mass elements;
however, this scenario has not been completely ruled out, and a lot of 
multidimensional works have
been devoted to it (e.g., \cite{GBW99,FHR07} and references therein)

\section{Accretion}


There is a variety of nuclear burning regimes on top of accreting white 
dwarfs, depending mainly on the accreted mass composition, the mass 
transfer (and thus the mass accretion) rate and the mass of the white dwarf. 
For a narrow range of accretion rates, burning can occur in steady state, 
so that matter is burned at the same rate as it is accreted. Large 
accretion rates are needed to fulfill this condition: 
$3.066 \times 10^{-7} \times$ (M/M$_\odot-0.5357)$ M$_\odot$/yr for 
H-accretion 
\cite{Nomoto07,SB07}, 
the exact value depending on the metallicity 
of the accreted matter; for He accretion, steady burning occurs for mass 
accretion rates roughly 10 times larger 
\cite{Fujimoto82a,Fujimoto82b,Iben82,JHI93}. 
It is worth mentioning that for larger accretion rates, the envelope 
expands and behaves like a red giant envelope, whereas for even larger rates 
the Eddington limit is reached, which inhibits further accretion (because 
radiation pressure force counteracts gravitational force). 

An interesting case of steady burning of hydrogen corresponds to the 
so-called supersoft X-ray sources (SSSs); these objects emit just in the 
``super soft'' X-ray energy range (i.e., below about 1 keV), and are 
interpreted as hot white dwarf photospheres, being hot because of steady 
burning of hydrogen) \cite{vdh92}. Provided that they 
have enough time to reach the Chandrasekhar mass, without any catastrophic 
mass-loss event, they are viable single degenerate scenarios for type Ia 
supernovae.

 
\begin{table}
\caption{Properties of models with weak H-shell flashes, for an  
initial white dwarf mass 1.2 M$_\odot$. Adapted from Jos\'e, Hernanz 
\& Isern \cite{JHI93}}
\begin{center}
\begin{tabular}{cccccc}
\hline
$\dot{M} (\rm M_\odot$/yr)  & $P_{rec}$ (yr) 
& $T_{max,H} (10^8$K) & $T_{min,H}$ ($10^7$K)
& $\overline{T_H}$ ($10^7$K) & $\overline{\rho_H}$ ($10^2$g/cm$^3$)\\
\hline
$2 \times 10^{-7}$      & 9
& 1.3                   & 3.4 
& 6.03            & 0.78\\
$10^{-7}$               & 22
& 1.4                   & 2.9 
& 4.36            & 1.20\\
$5 \times 10^{-8}$      & 50
& 1.4                   & 2.6 
& 3.63            & 1.58\\
$10^{-8}$               & 357
& 1.5                   & 2.0
& 2.69            & 2.57\\
$5 \times 10^{-9}$      & 832
& 1.6                   & 1.8
& 2.45            & 3.09\\
$10^{-9}$               & 5828
& 1.7                   & 1.5 
& 2.04            & 4.47\\
$5 \times 10^{-10}$     & 13691
& 1.7                   & 1.3
& 1.86            & 5.37\\
$2 \times 10^{-10}$     & 68640
& 1.9                   & 1.2
& 1.66            & 8.32\\
\hline
\end{tabular}
\label{tab:flashesH}
\end{center}
\end{table}

\begin{table}
\caption{Properties of models with weak He-shell flashes, for an  
initial white dwarf mass 1.2 M$_\odot$. Adapted from Jos\'e, Hernanz 
\& Isern \cite{JHI93}}
\begin{center}
\begin{tabular}{cccccc}
\hline
$\dot{M} (\rm M_\odot$/yr)   & $P_{rec}$ (yr) 
& $T_{max,He} $($10^8$K) & $T_{min,He}$ ($10^7$K)
& $\overline{T_{He}}$ ($10^8$K) & $\overline{\rho_{He}}$ ($10^4$g/cm$^3$)\\
\hline
$2 \times 10^{-6}$       & 199
&4.1                     & 11.5
& 1.55            & 0.85\\
$10^{-6}$                & 471
&4.3                     & 9.8
& 1.29            & 1.17\\
$5 \times 10^{-7}$       & 1124
&4.5                     & 8.7 
& 1.15            & 1.44\\
$10^{-7}$                & 9081
&5.0                     & 6.7 
& 0.98            & 2.34\\
$5 \times 10^{-8}$       & 22129
&5.2                     & 6.1 
& 0.93            & 2.82\\
\hline
\end{tabular}
\label{tab:flashesHe}
\end{center}
\end{table}

Whenever the accretion rate is smaller than the values corresponding to the 
steady burning regime, but larger than some critical values (explained below), 
accumulated matter reaches partial degeneracy and weak 
flashes occur. These flashes are not driving mass loss and they are 
strictly periodic. The general properties of these weak flashes are shown in 
Tables \ref{tab:flashesH} and \ref{tab:flashesHe}; these results were 
obtained by Jos\'e, Hernanz \& Isern \cite{JHI93}, with a semi-analytical 
model in the plane parallel approximation. It is shown that the 
recurrence period ($P_{rec}$) increases as the mass accretion rate 
diminishes, because 
a larger mass needs to be accreted to attain ignition conditions, both 
for H- and He-rich accretion. As a consequence of this increase, the 
intensity of the flashes increases as the accretion rate decreases; 
the reason is that  
a larger accreted mass implies higher degeneracy and thus a stronger flash 
\cite{JHI93} (as seen by the larger maximum shell temperatures, $T_{max}$, 
and larger difference between the maximum and minimum shell temperatures, 
$T_{max}-T_{min}$, obtained). 
The effect of the initial white dwarf mass goes in the same 
direction: the larger the initial white dwarf mass, the higher the pressure 
at the base of the accreted envelope (for a given accreted mass) and then the 
stronger the flash. Another indication of the strength of the flash is the 
ratio of the ``on phase'' duration to the recurrence period: the 
strongest flashes have small ratios, whereas the milder flashes have larger 
ratios (i.e., they stay in the ``on phase'' longer; see \cite{Iben82,JHI93} 
for details). The average values of shell temperatures $T_H$ and $T_{He}$ 
($\overline{T_{H}}$ and $\overline{T_{He}}$) and densities 
($\overline{\rho_{H}}$ and $\overline{\rho_{He}}$)
along the period are 
shown in Tables \ref{tab:flashesH} and \ref{tab:flashesHe}.

Up to now we have described direct accretion of H and He, but it is worth 
noticing that H burning (steadily or through weak flashes) leads to the 
accumulation of He. Therefore, one indirect way to accrete He is through
weak H flashes; in fact, double H-He flashes are expected for a range of
direct H accretion rates, e.g., those mentioned above as non problematic 
for H (see \cite{JHI93} and the complete hydrostatic simulations from  
Cassisi, Iben \& Tornambe and Piersanti et al. \cite{CIT98,PCIT99}). 
As shown in Tables \ref{tab:flashesH} and \ref{tab:flashesHe}, the 
typical period 
of single He flashes is about 400 times larger than that of H flashes, 
for a given (large) accretion rate;  
but the ratio $P_{rec,He}$/$P_{rec,H}$ changes when double flashes are 
considered. 
The existence of a top hydrogen shell influences the evolution 
of the bottom He layer, leading to a shorter recurrence period and, 
therefore, to a milder flash \cite{JHI93}. On the other hand, the 
accumulation of He as a consequence of H accretion can prevent the growth 
of the white dwarf mass up to the Chandrasekhar limit, since dynamical 
He flashes are probable \cite{CIT98,PCIT99}. More details, in the framework 
of hydrodynamical simulations, about the final flashing of He are given 
below, but just for direct He accretion.

Coming back to direct accretion, we have shown that the strength of the H or 
He flashes increases with decreasing accretion rate. For 
low enough accretion rates, flashes are not weak anymore, because 
accumulated matter can reach degenerate conditions and drive 
a hydrodynamic event, with potential mass loss. For H-rich mass accretion, 
nova explosions occur, leading to a large increase of visual luminosity and 
mass-loss at large velocities, from hundreds to thousands of km/s 
\cite{Her05,JH07}. 
For He accretion, there is a range of accretion rates for which there 
is a strong He flash (probably leading to a He detonation, i.e., supersonic 
burning propagation; see below). Therefore,  
weak flashes leading to a growing CO core towards the 
Chandrasekhar mass are not possible for this range of mass accretion rates. 

The critical accretion rates for surface degenerate burning and ensuing 
strong flashes are different for H and He. 
The lower limits for which weak H flashes occur are 
$\sim 10^{-9}-10^{-10}$ M$_\odot$/yr, the exact values depending on the 
initial mass and luminosity of the white dwarf and on the metallicity 
of the accreted matter \cite{MacDonald83}.
Concerning He, a strong flash is expected whenever the accretion 
rate ranges between $\sim 5\times 10^{-8}$ and $10^{-9}$ M$_\odot$/yr 
\cite{Nomoto82b,WTW86}, for any initial white dwarf mass. For smaller 
accretion rates, a strong flash is again predicted for white dwarf masses 
below $\sim 1.1$ M$_\odot$, but no He flash occurs if the white dwarf 
is more massive; alternatively, weak He flashes allowing for the increase 
of the white dwarf mass, as in the case of He accretion rates larger than 
$5\times 10^{-8}$ M$_\odot$/yr, would happen \cite{Nomoto82a}.

The further hydrodynamical evolution after a strong He flash on top of 
an accreting CO white dwarf has been subject to debate during many years, 
in the framework of both one dimensional and multidimensional simulations. 
It is worth noticing that in 1D simulations with spherical symmetry, ignition 
can only occur either at the center or simultaneously along a spherical 
shell. This instantaneous ignition throughout a shell seems unphysical, 
requiring multidimensional analyses; in fact, the initial stages of 
He ignition play a crucial role which deserves particular 
attention (see for instance \cite{GBW99} and \cite{FHR07} and references 
therein).
The original 1D works (e.g., \cite{Nomoto82b,WTW86}) tried first 
to elucidate if a He detonation wave propagating outwards formed as a 
consequence of the strong He flash.
 A related issue is to determine if the initial He detonation 
induces a carbon detonation at the core envelope interface, 
propagating inwards and leading to the so-called double 
detonation supernova \cite{Nomoto82b}. More recent studies devoted to 
CO white dwarfs with initial sub-Chandrasekhar mass accreting at the 
critical regime leading to external He detonation, 
indicated that C was not ignited at the core-envelope interface, 
so that there was not an inward C detonation wave; instead a compressional 
shock wave propagating inwards was responsible for the final C ignition,  
at the center or near it \cite{Livne90,WW94,Her97}.

As a summary, the outcome of He accretion in the critical range of 
accretion rates is not clear at all. Just as an illustration, 
let's mention 
Livne \& Glasner \cite{LG90}, who analyzed the consequences of an 
off-center He detonation: in 1D they did not obtain a double detonation, 
whereas in 2D they could neither confirm nor exclude it, but they mentioned 
that rarefaction effects not included in their analysis most probably would 
prevent it. In later numerical simulations \cite{LG91}, they concluded 
that a double detonation occurred for masses larger than 1.2 M$_\odot$, 
but was prevented in less massive white dwarfs.  
Fortunately, the gross nucleosynthesis predictions seem quite robust, and 
the predictions of 1D and 2D models agree quite well (see \cite{LA95}; a 
recent analysis of the nucleosynthesis constraints on the type of burning 
propagation can be found in \cite{GS07}). 


The final fate of all the scenarios described above, as well as the main  
characteristics of the ensuing type Ia explosion -whenever it occurs- are 
tightly related to the details of the burning propagation, which are 
reviewed by Roepke in this volume \cite{Roepke}. 
However, it is worth reminding that 
the presupernova evolution, leading to
the initial stages of the ignition, also plays a very important role on the 
further development of the explosion, whatever the scenario is. For instance, 
the presence of $^{14}$N in the accreted material is critical, since it 
allows the operation of the 
$^{14}$N(e$^-,\nu)^{14}$C($\alpha,\gamma)^{18}$O 
(NCO) reactions (above the density threshold $10^6$g cm$^{-3}$), 
which compete with the 3$\alpha$ ones at low temperatures 
and high densities \cite{Hashi86,WW94}. The main consequence is that the 
He layer is heated and ignition density becomes lower than in the models 
where the NCO reactions are not included; this could prevent the 
development of a strong He flash. However, Piersanti, Cassisi \& Tornambe 
\cite{PCT01} obtain that the NCO energy contribution is not able to 
keep the He shell hot enough to avoid strong electron degeneracy, and thus 
a He flash results, 
for realistic initial models of white dwarfs. The reason is that 
for low initial metallicities there is not enough $^{14}$N, whereas for 
solar or larger metallicities the densities attained are below the 
threshold for triggering the NCO reactions (notice that their models 
are for sub-Chandrasekhar mass white dwarfs, of 0.6 and 0.8 M$_\odot$).

Last but not least: an important factor which determines the effect of 
accretion onto white dwarfs is rotation. Yoon \& Langer \cite{YL94} 
studied the effect of the white dwarf spin-up  
caused by accretion of angular momentum, and the further rotationally 
induced chemical mixing; they deduced that the heating of the He envelope 
by friction related to differential rotation made it ignite under less 
degenerate conditions, and thus He detonation might be avoided (leading 
to a recurrent He nova explosion instead of a type Ia supernova).

\section{A particular scenario for type Ia supernovae: the RS Oph  
symbiotic recurrent nova}
There is a particular scenario for type Ia supernova that deserves 
attention, specially because there is a lot of recent and accurate 
observational information 
related with it: the symbiotic recurrent nova RS Oph. This phenomenon is 
related with H-accretion leading to nova outbursts, but with the 
peculiarity that the mass of the white dwarf is expected to increase 
(and not decrease) as a consequence of the explosion. Therefore, for 
an initial white dwarf massive enough, the Chandrasekhar mass can be 
reached in a reasonable amount of time.

RS Oph has undergone various recorded nova outbursts (that's why it 
belongs to the recurrent nova class), with a recurrence period of 21 years, 
the last one in 2006. It was observed 
at practically all wavelengths, spanning the range from radio to 
hard X-rays. Its orbital period is 456 days and the companion of the 
white dwarf is a red giant star, with a wind. This scenario is different 
from the scenario of ``standard'' classical novae, where the companion is a 
main sequence star and the periods - and thus orbital separations - are much 
smaller. As a consequence, the accretion rate in RS Oph is much larger and 
the time needed to accrete the critical mass to develop a nova outburst much 
shorter (some decades, instead of the typical recurrence periods 
$\sim 10^4-10^5$ years of classical novae). According to what was mentioned in 
the previous section, for large accretion rates, e.g. 
$10^{-6}-10^{-7}$ M$_\odot$/yr, there is not degenerate H-burning and 
thus no nova explosion is expected; however, if the initial mass of 
the white dwarf is large enough (very 
close to the Chandrasekhar mass), the explosion occurs even for such large 
accretion rates. Recent models by Hernanz \& Jos\'e \cite{HJ08}, have 
successfully reproduced the global properties of RS Oph, with white dwarf 
initial masses 1.35 and 1.38 M$_\odot$ and accretion rates (of matter 
with solar composition, from the red giant companion) about 
$2 \times 10^{-7}-10^{-8}$ M$_\odot$/yr. An important result is that 
the accreted masses 
are larger than the ejected ones, so that the mass of the white dwarf 
grows towards the Chandrasekhar mass; it increases by $10^{-6}$ M$_\odot$, 
typically, after each eruption. The corresponding time needed to 
reach the mass limit, 
i.e., to reach explosion conditions, ranges between 3 and $7 \times 10^5$ 
years (for an accretion rate of $2 \times 10^{-7}$ M$_\odot$/yr). 

Therefore, RS Oph and its relatives (just a few -less than 5- in the
Galaxy) are 
potential type Ia candidates, in the framework of the single degenerate 
scenario. But a problem remains with this scenario: a white dwarf of such 
large initial mass should be made of ONe, and not of CO, according to 
standard stellar evolution; then it would collapse and not explode as a type 
Ia supernova. There is not a clear way to solve this conflict, unless there 
was some previous epoch of steady burning or very weak flashes, allowing 
for the growth of the mass from $\sim 1.1$ M$_\odot$ (the maximum mass 
of a CO core ''at birth'') to the values of the white dwarf mass required 
to explain the short recurrence period ($\sim 1.35-1.38$ M$_\odot$). An 
additional problem with this scenario is how to get rid of the hydrogen, 
which has been observed in large amounts in the expanding ejecta. But this 
is a long standing problem of the single-degenerate scenario, which, however 
seems not to be in contradiction with the observations (see section 1), 
according to some recent papers.


\acknowledgments{This research has been funded by the spanish MEC grants 
ESP2007-61593 and AYA2007-66256, by the catalan AGAUR grant 2005-SGR00378 
and by FEDER funds.}

\end{document}